\documentclass[a4paper,fleqn,usenatbib]{mnras}

\usepackage{newtxtext,newtxmath}

\usepackage[T1]{fontenc}
\usepackage{ae,aecompl}

\usepackage{graphicx}	
\usepackage{amsmath}	
\usepackage{amssymb}	

\title[magnetic background spectrum]{Solar wind magnetic field background spectrum from fluid to kinetic scales}
\author[R. Bruno et al.]{Roberto Bruno$^{1}$\thanks{Contact e-mail: \href{mailto:roberto.bruno@iaps.inaf.it}{roberto.bruno@iaps.inaf.it}}, Daniele Telloni$^{2}$, Danilo DeIure$^{3}$ and Ermanno Pietropaolo$^{3}$
\\
$^{1}$National Institute for Astrophysics, Institute for Space Astrophysics and Planetology, Via del Fosso del Cavaliere 100, 00133 Roma, Italy\\
$^{2}$National Institute for Astrophysics, Astrophysical Observatory of Torino, Via Osservatorio 20, 10025 Pino Torinese, Italy\\
$^{3}$University of L'Aquila, Department of Physical and Chemical Sciences, L'Aquila, Italy
}
\pubyear{2017}
\begin{document}
\label{firstpage}
\pagerange{\pageref{firstpage}--\pageref{lastpage}}
\maketitle
\begin{abstract}
The solar wind is highly structured in fast and slow flows. These two dynamical regimes remarkably differ not only for the average values of magnetic field and plasma parameters but also for the type of fluctuations they transport. Fast wind is characterized by large amplitude, incompressible fluctuations, mainly Alfv\'{e}nic, slow wind is generally populated by smaller amplitude and less Alfv\'{e}nic fluctuations, mainly compressive.
The typical corotating fast stream is characterized by a stream interface, a fast wind region and a slower rarefaction region formed by the trailing expansion edge of the stream.
Moving {between these two regions}, from faster to slower wind,  we observe the following behavior:
a) the power level of magnetic fluctuations within the inertial range largely decreases, keeping the typical Kolmogorov scaling;
b) at proton scales,  for about one decade right beyond the high frequency break, the spectral index becomes flatter and flatter towards a value around -2.7;
c) at higher frequencies, before the electron scales, the spectral index remains around -2.7 and,  {based on suitable observations available for $4$ corotating streams}, the power level does not change, irrespective of the flow speed.
All these spectral features, characteristic of high speed streams, suggest the existence of a sort of magnetic field background spectrum. This spectrum would be common to both faster and slower wind but, any time the observer would  cross the inner part of a fluxtube channeling the faster wind into the interplanetary space, a turbulent and large amplitude Alfv\'{e}nic spectrum would be superposed to it.

\end{abstract}
\section{Introduction}
\label{sec:introduction}
Typical corotating high speed streams (HSS' hereafter), i.e. those streams coming from the equatorial extension of polar coronal holes, are characterized by different regions. Within these regions, extending on daily scales, field and plasma parameters assume different average values and fluctuations have a different character. The first region is the stream interface (SI), located between fast and slow stream, right where the dynamical interaction between the two flows is stronger  {\citep{schwenn1990, brunocarbone2016}}. The SI is characterized by high level of pressure, both kinetic and magnetic, high temperature and low Alfv\'{e}nicity. This region is followed by the trailing edge (TE), which is characterized by the highest wind speed and temperature, the lowest compressibility, and the highest level of Alfv\'{e}nicity. The TE is then followed by a wind which progressively slows down and becomes cooler. This region was firstly named rarefaction region by \cite{hundhausen1972}. Interesting enough, these two portions of the stream are separated by a very narrow region across which Alfv\'{e}nicity changes abruptly, from high to low  {\citep{bruno2015}}. Most of the time, the rarefaction region is followed by the heliospheric current sheet crossing characterized by low Alfv\'{e}nicity, low temperature but higher field and plasma compressibility. Thus, moving from fast to slow wind one observes a different turbulence, not only relatively to its power level but also to its Alfv\'{e}nic character although the spectral slope at fluid scales clearly shows a persistent Kolmogorov-like scaling.
In addition, besides the above-cited differences within the fluid regime, there are also clear differences at proton kinetic scales. \cite{bruno2014b}, investigating the behavior of the spectral slope at proton scales, up to frequencies of a few Hz, beyond the high-frequency break separating fluid from kinetic scales, recorded a remarkable variability of the spectral index \citep{smith2006,sahraoui2010} within the following frequency decade or so. The steepest spectra corresponded to the trailing edges of fast streams while the flattest ones were found within the subsequent slow wind regions. The same authors found an empirical relationship between the power associated to the inertial range and the spectral slope observed within this narrow region which allowed to estimate that this slope approaches the Kolmogorov scaling within the slowest wind and reaches a limiting value of roughly $−4.4$ within the fast wind. In addition, \cite{bruno2014b} suggested the possible role played also by Alfv\'{e}nicity in this spectral dependence. Recent theoretical results seem to move in this direction attributing a relevant role to Alfv\'{e}nic imbalance within the inertial range \citep{voitenko2016}.

The location of the frequency break and the nature of the fluctuations within this restricted frequency range was addressed by several authors within the recent literature on this topic (see reviews by \cite{alexandrova2013, brunocarbone2016}). Although turbulence phenomenology would limit the role played by parallel wavevectors $k_\parallel$, \cite{bruno2014a} firstly found that the cyclotron resonant wavenumber showed the best agreement with the location of the break compared to results obtained for the ion inertial length and the Larmor radius. Successively, \cite{chen2014} found similar results but did not consider a possible role of $k_\parallel$ because it was not consistent with the observed anisotropy of turbulence \citep{horbury2008}. Thus, this problem is still open and these observations expect to have a theoretical explanation.
\cite{telloni2015} analysed the nature of the magnetic fluctuations within this limited frequency range beyond the spectral break and found that their character was compatible with left-hand, outward propagating ion cyclotron waves and right-hand kinetic Alfv\'{e}n waves, confirming previous findings by authors like \cite{he2011, he2012a, he2012b, podesta2011}.

In particular, \cite{bruno2015} found that these two populations of fluctuations experienced a different fate within different regions of the velocity stream. Both fluctuations become indeed weaker and weaker when analysing slower and slower wind, and the left-handed population would be the first one to disappear.
Another interesting feature of the analysed spectra, which stimulated the present work, was represented by the results shown in Figure $4$ of \cite{bruno2014b}. Those spectra refer to different time intervals chosen along the speed profile, from fast to slow wind. The low frequency range and the high frequency range were obtained by Fourier transforming WIND data from the fluxgate magnetometer and THEMIS-C data from the searchcoil magnetometer, respectively. While within the inertial range the power density level greatly differs from fast to slow wind, as expected, no evident differences can be observed within the high frequency range above 1Hz, all the spectra collapse on top to each other showing a spectral slope that \cite{bruno2014b} estimated to be roughly $-2.5$.

\begin{table}
\caption{Time intervals referring to the 4 high-speed streams observed
with CLUSTER and THEMIS, along with the corresponding average wind speed.}
\label{tab:high-speed_streams}
\begin{center}
\begin{tabular}{ccccc}
\hline
\hline
\# & Year & Time interval & s/c & $\langle$V$\rangle$ \\
& & DoY:hh & & km s$^{-1}$ \\
\hline
{1} & {2007} & 72:06 - 73:09 & CLUSTER-1 & 671 \\
                   &                       & 74:15 - 75:18 & CLUSTER-1 & 612 \\
                   &                       & 79:09 - 80:12 & CLUSTER-1 & 296 \\
\hline
{2} & {2011} & 36:18 - 37:01 & THEMIS-C & 580 \\
                   &                       & 37:18 - 38:01 & THEMIS-C & 527 \\
                   &                       & 39:08 - 39:21 & THEMIS-C & 422 \\
                   &                       & 40:18 - 41:01 & THEMIS-C & 323 \\
\hline
{3} & {2011} & 149:15 - 150:00 & THEMIS-C & 680 \\
                   &                       & 152:15 - 153:00 & THEMIS-C & 514 \\
                   &                       & 153:15 - 154:00 & THEMIS-C & 447 \\
                   &                       & 154:15 - 155:00 & THEMIS-C & 399 \\
\hline
{4} & {2011} & 174:15 - 175:00 & THEMIS-B & 605 \\
                   &                       & 178:15 - 179:02 & THEMIS-C & 408 \\
                   &                       & 179:15 - 180:00 & THEMIS-B & 389 \\
\hline
\hline
\end{tabular}
\end{center}
\end{table}

A similar study was already produced by \cite{alexandrova2009} who showed that spectra measured under different plasma conditions have a similar shape concluding that there is a sort of universality of solar wind turbulent spectrum from MHD to electron scales. However, this universal spectrum obtained by \cite{alexandrova2009} was the result of a normalization of the spectral power density adopted in order to collapse all the spectra on top of each  other to highlight a remarkable similarity in the spectra. The main difference with the analysis by \cite{bruno2014b} is that these authors did not apply any normalization and the collapsing of the spectra within the high frequency range came out naturally. Moreover, another major difference was in the choice of the intervals. In \cite{alexandrova2009} most of the intervals belonged to compressive regions while, in our analysis, they were avoided since we aimed to show the spectral evolution from fast to slow wind within typical corotating  HSS.

Thus, the main goal of the present work is that of checking whether the results highlighted in Figure 4 by \cite{bruno2014b} are fortuitous or represent the canonical situation within corotating  HSS.

\section{Data analysis and results}
\label{sec:data_analysis_results}

In order to perform this kind of analysis it is necessary to have observations covering a wide range of time scales. Since we aim to study the magnetic field power density spectrum between fluid and proton kinetic scales we need to cover time scales ranging from several tens of minutes to tens of hertz. Such a wide dynamical range is generally covered by two different kind of magnetometers, one for the low frequency or DC field and another one for the high frequency or AC field. Two common kinds of magnetometers adopted onboard magnetospheric and heliospheric s/c are flux-gate and the search-coil instruments for DC and AC field, respectively \citep{pfaff1998}.
For our purposes, it is necessary that both instruments operate within the same time interval although we are aware that the search-coil data are available only for a much shorter time with respect to the fluxgate. Moreover, the present study limits the observations to the solely corotating high-speed streams, not influenced by magnetospheric up-stream waves or solar wind transients, and the following low-speed incompressive region. We paid particular attention in avoiding slow wind compressive regions generated by the dynamical interaction between fast and slow wind, i.e. stream-stream interface, or belonging to the heliospheric current sheet crossing rather than to the slow wind tail of the corotating high-speed stream.

\begin{figure}
	\includegraphics[width=7.5cm]{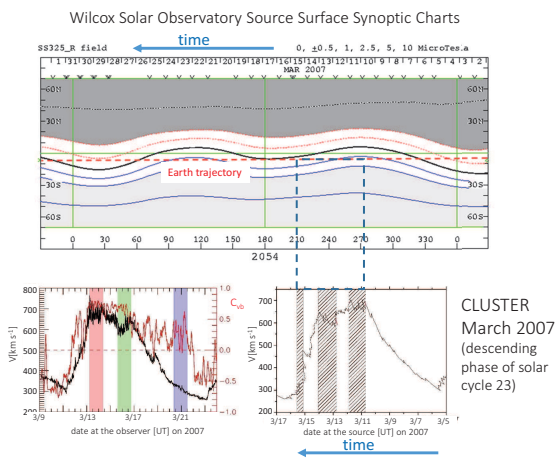}
	\caption{Top panel: source surface synoptic chart from Wilcox Observatory, evaluated at $3.25$ solar radii, for solar rotation $\# 2054$. Blue contour lines and light grey shading show positive magnetic polarity regions while dark grey area and dotted red lines refer to the opposite polarity. The black solid line shows the location of the neutral line. The dashed red line shows the Earth's trajectory as projected onto the Sun. Time runs from right to left.
Lower left-hand-side panel: speed profile (black colour line) of the stream of interest between March $11$ and $23$ and Alfv\'{e}nic correlation (red colour line). The three coloured vertical bars highlight the selected three time intervals. Lower right-hand-side panel: profile of the same stream projected back to the Sun using $1$ hr solar wind speed averages.
}
	\label{fig_01}

\end{figure}

Among all the past and current space missions probing the solar wind, the only two allowing this kind of study, with available data freely accessible on the web, are THEMIS and CLUSTER, two constellations of $5$ and $4$ flight-formation s/c, respectively.
However, both missions were primarily designed for studying the Earth's magnetosphere and only for relatively short periods of time CLUSTER and THEMIS are/were immersed in the solar wind. Unfortunately, it was possible to use THEMIS observations only during the relatively short period of time when THEMIS B and C s/c were transferring from terrestrial to lunar orbit. On the other hand, Cluster did not offer a much better situation relatively to the choice of appropriate time intervals. Indeed, the average Cluster orbital period of about $2$ days and the average short duration of a high-speed stream of about $1$ week makes quite low the probability that the s/c probes both the trailing edge and the rarefaction region of the stream during consecutive passages at the apogee. Moreover, the apogee is in the solar wind only for a limited period of time during the year. As  a consequence, all the data analysis requirements on the choice of the time intervals and the orbital features of the above two missions made the data selection extremely difficult and we were able to select only $4$  HSS' from Cluster and THEMIS B and C s/c, listed in Table \ref{tab:high-speed_streams}.

The CLUSTER data repository provides magnetic field vector measurements from the FGM (Flux Gate Magnetometer) instrument with a cadence of $0.2$ sec. The same repository provides also directly power density spectra computed from SCM (Search Coil Magnetometer) data sampled between $0.1$ Hz to $4$ kHz. The THEMIS data repository
provides $4$ and $128$ Hz resolution measurements from fluxgate and search-coil magnetometers, respectively.

Particular attention has to be paid when analyzing FGM data at frequencies larger than a few Hz since this frequency range can be affected by noise due to the digitization process \citep{bennett1948, russell1972}. As reported by \citet{bruno2014b}, the spectral level due to the digitization noise tends to flatten out the power spectrum of FGM data for frequencies above roughly $2-3$ Hz. Hence, in the present analysis, the study of the power spectra obtained from both CLUSTER and THEMIS fluxgate measurements is limited to frequencies lower than $2$ Hz.

As discussed in \citet{alexandrova2013}, CLUSTER search-coil measurements are severely affected by instrumental noise above $100$ Hz. Hence, in the present work, the power spectra obtained from SCM data are displayed up to $64$ Hz, to avoid noise problems and to match the Nyquist frequency of THEMIS search-coil measurements.
Moreover, CLUSTER search-coil data are provided already filtered at $0.6$ Hz and data below this value are not relevant and should not be used.
In our case, we adopted a conservative choice showing spectra only above $2$ Hz which, on the other hand, is the high frequency limit of spectra built from FGM data.

It is worth reminding that the THEMIS probes spin with a frequency of $1/3$ Hz. This leads to an artificial large amplitude periodic signal at the spin frequency in the SCM measurements (see \citet{lecontel2008} for a complete treatment of spurious signals and noise in THEMIS data).
To remove the spin modulation effects from the spectral analysis, THEMIS search-coil data are provided already high-pass filtered at a frequency slightly higher than the spin frequency. However, in the present analysis, the power spectra obtained from THEMIS search-coil measurements are shown only for frequencies above $2$ Hz, which, on the other hand, is the highest frequency of the spectra we show from the fluxgate data.

THEMIS search-coil data are furthermore affected by two different sources of spurious noise: large amplitude spikes at twice the spin frequency ($1/3$ Hz), and at $8$ and $32$ Hz, \citep{lecontel2008}.  Their harmonics are indeed found to dominate the power spectrum at higher frequencies. However, since both types of spurious noises are phase locked (with the spin frequency and with the onboard $1$ second instrument clock, respectively) and are quite constant in amplitude with time, it is possible to reduce their level by means of the Superposed Epoch Analysis (SEA) as described in \citet{lecontel2008}.

We applied this SEA technique to THEMIS search-coil data before the spectral analysis for all the time intervals we chose. However, as reported in \citet{lecontel2008}, this technique, very effective in reducing phase-locked noise, is not equally effective in reducing noise that is not phase-locked and whose effects are still present in the spectra after SEA filtering, as we will comment in the following.

We compute Power Spectral Densities (PSD) of magnetic field data from fluxgate and search-coil magnetometers onboard CLUSTER and THEMIS s/c from the trace of the spectral matrix using the Morlet Wavelet Transform. In particular, the wavelet PSD shown in this paper corresponds to the global wavelet spectrum (that is, the time-averaged wavelet spectrum over the whole time interval), which provides an unbiased and consistent estimation of the true power spectrum of a time series \citep{percival1995}.


\begin{figure}
	\centering
	\includegraphics[width=7.7cm]{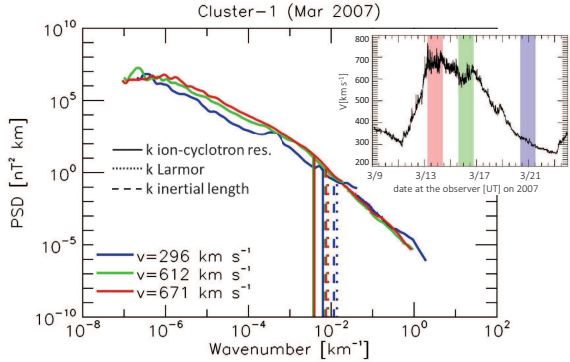}
	\caption{Spectral analysis of the selected time intervals as indicated in the inset plot by different coloured bars. Vertical lines refer to the position of the wavenumbers corresponding to the
the proton inertial length $\kappa_i$, the proton Larmor radius $\kappa_L$ and the smallest wavenumber corresponding to the resonance condition for parallel propagating Alfv\'{e}n waves $\kappa_r$.}
	\label{fig_02}

\end{figure}

The first  HSS we analysed was observed by Cluster in mid March 2007, during the descending phase of solar activity cycle $23$.
The top panel of Fig. \ref{fig_01} is a source surface synoptic chart from Wilcox Observatory. This chart, which covers solar rotation $\# 2054$, shows contour levels of magnetic field intensity evaluated at $3.25$ solar radii \citep{hoeksema1983}.
Blue contour lines and light grey shading show positive magnetic polarity regions while dark grey area and dotted red lines refer to the opposite polarity. The black solid line shows the location of the neutral line, or heliomagnetic equator, that smoothly runs between the two opposite unipolar magnetic regions, characteristic of low solar activity. The dashed red line shows the Earth's trajectory as projected onto the Sun. In this kind of map the time runs from right to left.
The lower left-hand-side panel shows the speed profile of the stream of interest between March $11$ and $23$. The data shown in this panel are $1$ min plasma averages taken from Wind because the vantage location of this s/c at the Lagrangian point L1 allows to monitor the solar wind continuously. This is a typical high-speed stream with a bulk velocity ranging roughly between $750$ and $300$ km/s.
The three coloured vertical bars highlight the three time intervals that we could select due to the restrictions imposed by the orbit of Cluster s/c and the availability of the data.
In the lower right-hand-side panel we report the profile of the same stream projected back to the Sun using $1$ hr averages of solar wind speed. Although relatively low time resolution, $1$ hr is largely appropriate to back project onto the solar surface in-situ measurements. In this panel, accordingly with the top panel, time is running from right to left.
Due to the different values of the wind speed at the beginning and at the end of each timer interval, the temporal width of the regions mapped onto the Sun differs from the original duration of the time intervals at 1 AU (see left-hand-side panel). This back-projection shows that Cluster, during the selected time intervals, was immersed in the equatorial extension of a southern polar coronal hole.

\begin{figure}
	\includegraphics[width=7.5cm]{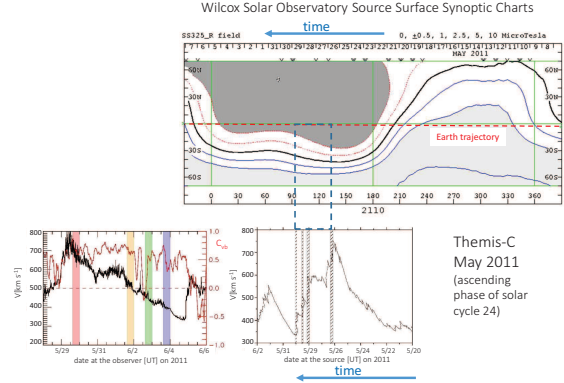}
	\caption{ HSS observed by THEMIS-C in May 2011 is shown in the same format of Fig. \ref{fig_02}}
	\label{fig_03}

\end{figure}

The highest speed is recorded when the s/c is farther in latitude with respect to the local heliomagnetic equator and the lowest speed is recorded when the s/c is approaching the smallest heliolatitude \citep{bruno1986}, i.e. the border of the coronal hole.
The presence of spikes visible between the beginning of March 13 and the end of March 16 (left-hand-side panel) reveals the Alfv\'{e}nic nature of those fluctuations \citep{matteini2014}. This consideration is indeed strengthened by the corresponding values assumed by the red curve that runs through the panel. This curve shows the value of $R_{VB}$ that represents the Alfv\'{e}nic correlation coefficient. This parameter takes into account not only the degree of correlation between magnetic and kinetic fluctuations but also the degree of equipartition between the two since $R_{VB}=\sigma_C/\sqrt{1-{\sigma_R}^2}$ where $\sigma_C/$ and $\sigma_R$ are the normalized cross-helicity and residual energy \citep{bavassano1998}. In particular, the Alfv\'{e}nicity curve shows a rapid increase through the velocity enhancement around March 13 and an abrupt decrease at the beginning of day 17 when we enter the rarefaction region of the stream (see also \cite{telloni2016} for similar events). This is a common characteristic of corotating  HSS' and the abrupt decrease in Alfv\'{e}nicity might be related to particular physical conditions developing during the expansion or inherent of the peripheral regions of the coronal hole. Moreover, the Alfv\'{e}nic character of the fluctuations continues to decrease as we get closer to the typical slow wind around day $22$ of March.

In Fig. \ref{fig_02} we show the results of our spectral analysis relative to the selected time intervals within the stream observed by Cluster.
To make it easier for the reader to follow our results, we show again the relative location of the selected time intervals within the stream in the inset plot. Different colours refer to the different spectra shown in the main panel. For sake of completeness, we show also the position of the wavenumbers corresponding to the proton inertial length $\kappa_i=\omega_{p}/c=\Omega_{p}/v_A$ and the proton Larmor radius $\kappa_L=\Omega_{p}/v_{th}$, where $c$ is the speed of light, $\omega_{p}$ is the proton plasma frequency and $\Omega_{p}$ is the ion-cyclotron frequency.

\begin{figure}
	\centering
	\includegraphics[width=7.7cm]{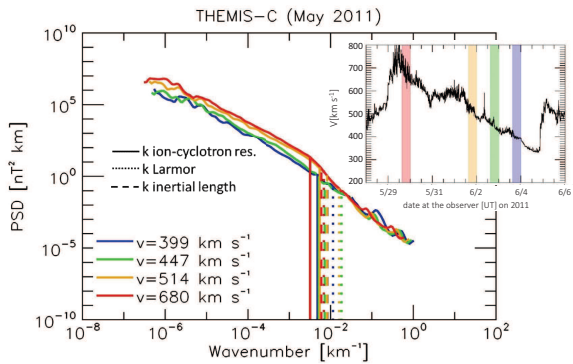}
	\caption{Power density spectra of the time intervals indicated in Fig. \ref{fig_03} in the same format adopted for Fig. \ref{fig_02}.}
	\label{fig_04}

\end{figure}

Moreover, we also show the position of the minimum wavenumber corresponding to the resonance condition for parallel propagating Alfv\'{e}n waves $\kappa_r=\Omega_p/(v_A+v_{th})$ since, as already shown by \cite{bruno2014a}, this wavenumber gives the best match with the observed location of the spectral break.

Moving from red to blue colour is equivalent to move from fast to slow wind. All the three spectra show the two different scalings  {characteristic} of the fluid and kinetic range, as we will discuss in the following. In addition, as expected, the transition region between these two regimes \citep{bruno2014b} is particularly clear for the fast wind samples and it is well indicated by the location of the estimated value of $\kappa_r$. The different spectral index is not the only feature that characterizes these two regimes. As a matter of fact, while the inertial range shows large differences in the power associated to magnetic fluctuations of fast and slow wind, the level of the fluctuations within the kinetic regime does not change appreciably, confirming previous results obtained by \cite{bruno2014b}.

As representative of the remaining three  HSS' observed by THEMIS we will show and discuss in detail only one of them since the results are very similar. In any case, we will show all the spectra from all the analysed time intervals together in summary plots that will be discussed later in the paper.

This corotating  HSS is shown in Fig. \ref{fig_03}, in the same format of Fig. \ref{fig_02}. These observations were taken by THEMIS-C about four years after the corotating stream observed by Cluster. This time interval falls within the ascending phase of solar cycle $24$ while the previous one was recorded during the descending phase of the previous cycle. The magnetic configuration at the Sun has changed from the time interval previously discussed, and the magnetic equator is much more warped than before, anticipating the general magnetic field inversion that would take place around the maximum of the same cycle, a few years later. For this stream there were four possible time intervals to analyse between fast and slow wind, as highlighted by the coloured stripes across the lower left-hand-side panel. The projected location of the s/c onto the Sun (lower right-hand-side panel) shows that THEMIS-C orbit was immersed in the equatorial extension of a northern polar coronal hole.

\begin{figure}
	\centering
	\includegraphics[width=6.2cm]{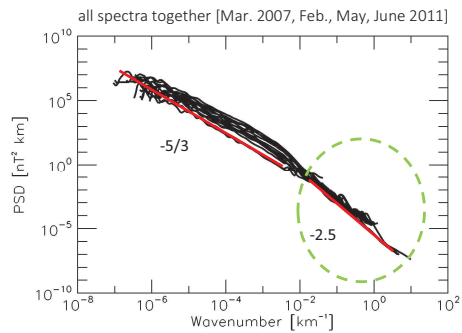}
	\caption{All the spectra from all the $14$ available time intervals listed in Table \ref{tab:high-speed_streams}. The red coloured segments highlight the approximate position of the magnetic background spectrum. }
	\label{fig_05}

\end{figure}

Power density spectra related to these intervals are shown in Fig. \ref{fig_04} in the same format adopted for Fig. \ref{fig_02}.
As already observed for the latter, spectra of fluid and kinetic scales behave differently when moving from fast to slow wind samples. To the large variability in the inertial range corresponds a rather stable spectrum at proton scales although the noise that could not be filtered, as previously discussed, causes the bumps that we notice at kinetic scales.  Moreover, also in this case, the break location is in better agreement with the ion-cyclotron resonance prediction than with the wavenumbers corresponding to the inertial length or Larmor radius.

Specific results from the other two time intervals analysed in this work will be omitted for sake of brevity, as already stated before, but all the spectra from all the $14$ available time intervals are reported in Fig. \ref{fig_05}. Although Cluster and THEMIS intervals belong to different phases of the solar cycle, the level of the kinetic part of their respective spectra, encircled by the green dashed line, remains remarkably constant in time while the level of the inertial range varies from fast to slow wind as expected. The spectral slopes marked by the red lines are shown just to drive the eye since the main goal of this paper is to show the different spectral variability between fluid and kinetic scales. As matter of fact, while the scaling at kinetic scales estimated from CLUSTER time intervals is around $-2.6$ the scaling estimated from all the THEMIS intervals is around $-2.3$ due to the unavoidable effect of the residual noise not phase locked and thus not filtered by the SEA technique.

The above results introduce some constraints on the spectrum. We have to consider the following points:

a) the border between fluid and kinetic scales is reasonably well identified by the wavenumber associated with the ion-cyclotron resonance mechanism \citep{bruno2014a};

b) the spectral slope in the inertial range is well approximated by the Kolmogorov's scaling;

c) fluid and kinetic spectral branches must be connected.

It follows that there should be a lower threshold for the spectrum of magnetic fluctuations and it can be graphically identified by the red segments in Fig. \ref{fig_05}; a sort of background spectrum  {characteristic} of corotating  HSS.

A further property of interplanetary magnetic field spectra can be used to infer what should be the highest possible level of spectral density associated with turbulence fluctuations characterizing the inertial range.
As a matter of fact, if we consider the following points:

a) the location of the break is between $3$ and $4\times10^{-3} \mathrm{km^{-1}}$ \citep{bruno2014a};

b) the maximum spectral slope within the transition region is roughly  $-4.4$ \citep{bruno2014b};

c) the width of this transition region is less than $1$ decade, as inferred from previous studies \citep{bruno2014b};

\begin{figure}
	\centering
	\includegraphics[width=6.2cm]{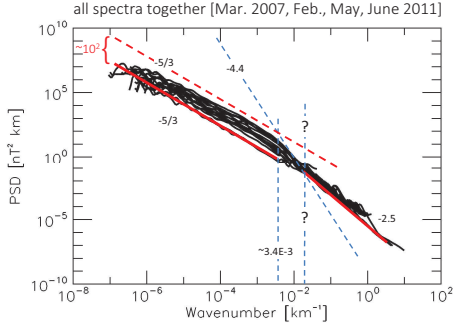}
	\caption{Geometrical construction to estimate the upper bound of the magnetic power spectral density within  HSS. See text for details.}
	\label{fig_06}

\end{figure}

we can draw the upper bound for magnetic fluctuations within the inertial range as shown by the dashed red line in Fig. \ref{fig_06}. At this point, we can estimate the maximum width of the interval of variability for the power spectral density between slow and fast wind. This interval results to be of the order of $2$ decades, as indicated in the plot. Obviously, there are uncertainties associated with this rough estimate but the expected effect cannot produce differences by orders of magnitude.

\begin{figure}
	\centering
	\includegraphics[width=8.3cm]{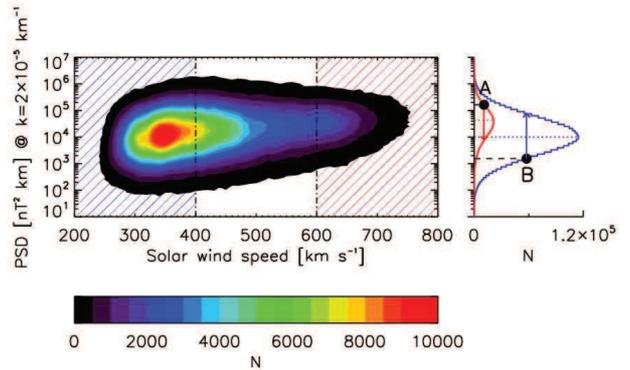}
    \caption{Main plot: 2D histogram of the total magnetic PSD, for a scale of $2\times10^{-5} \mathrm{km^{-1}}$, as a function of wind speed.
Right-hand-side graph: histograms built for slow ($< 400$ km/s) and fast ($> 600$ km/s) wind, respectively. The difference between points "A" and "B" is an estimate of the range of variability of the PSD from slow to fast wind.}
	\label{fig_07}
\end{figure}

In order to make this prediction more robust from a statistical point of view, we analyzed $12$ years of $1$-min magnetic field measurements of the WIND s/c from 2005 to 2016.
Based on the Morlet wavelet transform, we computed the PSD of magnetic field fluctuations at a single time scale well within the inertial range, namely $2\times10^{-5} \mathrm{km^{-1}}$.
To make this statistical analysis the most compatible with the spectral analysis discussed so far, we eliminated from the WIND dataset all the intervals corresponding to stream interfaces and to interplanetary coronal mass ejections, as catalogued by Ian Richardson and Hilary Cane since January 1996 \citep{richardson2010}. In this way, we focused  on corotating high-speed streams, mostly. The 2D histogram of Fig. \ref{fig_07} shows the distribution of the PSD as a function of wind speed, from the slowest to the fastest wind.

 This distribution appears to be asymmetric being characterised by a wider range of variability within the slow wind. As expected, larger PSD is preferentially associated with faster wind. The $2$ distributions shown in the right-hand-side graph are two histograms built for slow (i.e. $< 400$ km/s) and fast (i.e. $> 600$ km/s) wind, respectively. They show the range of variability for the PSD on a scale of $2\times10^{-5} \mathrm{km^{-1}}$ within these velocity intervals. The peaks of the distributions are well separated by slightly less than $1$ decade. However, if we take into account the variability of the PSD we have to consider the difference between points "A" and "B" which have been set at half maximum of their respective distributions. In doing so we end up with a total variability of the order of $2$ decades, similar to the estimate obtained from Fig. \ref{fig_06}.

 It is interesting to notice the rather flat top of the distribution extending from slow to fast wind. This represents a sort of limiting value for the PSD for the scale studied here. In the framework of a non-linear turbulent cascade experienced by these fluctuations, this limiting value should depend on the power associated with the injection scales, which, in turn, should not depend on the wind speed either.
 In particular, it is possible, although a dedicated analysis would be needed before any conclusion, that the larger variability within the slow wind is probably due to the presence of those slow wind intervals characterized by large amplitude Alfv\'{e}nic fluctuations recently studied by \citet{damicis2015}. As a matter of fact, these authors found that the level of Alfv\'{e}nic fluctuations within the Alfv\'{e}nic slow wind are of the same order of amplitude of those characteristic of the fast Alfv\'{e}nic wind.

A final consideration would suggest that, using the radial dependencies found by \cite{bavassano1982} it would be possible to predict the upper level of the inertial range spectrum that future space missions like Parker Solar Probe and Solar Orbiter might observe in the inner heliosphere. Using the radial dependence $R^{-4.2}$ found by these authors for the inertial range, we would expect to find an upper limit at $0.29$ AU about $1.8\times10^2$ times higher than the upper limit found at $1$ AU shown in Fig. \ref{fig_06}.

\section{Summary and discussion}
\label{sec:summary_discussion}
Power spectral density of interplanetary magnetic field fluctuations shows an unpredicted behavior at kinetic scales. Irrespective of what happens at fluid scales, the power spectral density level of kinetic scales remains largely unchanged when analysing either fast or slow wind data taken within corotating  HSS'.

Because of this and because of the roughly fixed location of the break point separating fluid from kinetic scales, we can establish, for the slow wind, a sort of background spectrum extending from fluid to proton scales. The power spectral level of this spectrum remains frozen when we look at different streams even if they are largely separated in time, i.e. coming from different source regions. Slower solar wind comes from the peripheral areas of the equatorial extension of a polar coronal hole and is relatively poor in Alfv\'{e}nic fluctuations. The present literature suggests that this spectrum is likely due to the non-propagating background magnetic structure advected by the wind \citep{tumarsch1995, brunocarbone2016}.
The Alfv\'{e}nic spectrum, which characterizes the fastest wind found in the trailing edge of the streams, is then superimposed onto the background spectrum and can vary in power from one stream to the other.
Alfv\'{e}nic magnetic fluctuations, being much larger than those due to the advected magnetic structure, completely cover the  {background} spectrum.
However, it is still unclear what determines the spectral differences between these two contiguous regions of the HSS. Is it due to differences inherent the nature of the source regions or is it due to some mechanism that acts differently during the expansion? A dedicated study to this topic is highly needed and future observations by Parker Solar Probe and Solar Orbiter will test these different possibilities.
In any case, also for the Alfv\'{e}nic spectrum there seems to be a threshold which, in this case, limits the maximum amplitude of the fluctuations, as expected for any physical phenomenon. The fact that the level of the kinetic spectrum does not depend on the level of the inertial spectrum suggests that the magnetic energy associated with the Alfv\'{e}nic fluctuations is somehow transferred to particles, within the spectral transition region, to modify their plasma kinetics. Several mechanisms, invoking wave particle interaction to replace collisions in a non-collisional fluid like the solar wind, or invoking  the role of strong field fluctuations within current sheets generated by turbulence reconnection processes able to energize particles, have been proposed (see reviews by \cite{alexandrova2013}, \cite{marsch2006} and \cite{brunocarbone2016} and references therein).
Unfortunately, we do not yet have observations of the particles velocity distribution function at kinetic scales in the solar wind to better understand which are the main physical mechanisms at work. However, we are confident that ESA-Solar Orbiter, NASA-Parker Solar Probe and possibly ESA-THOR, will provide all the necessary observations to shed light on this fundamental question.

\section*{Acknowledgements}
One of the Authors (DT) is financially supported by the Italian Space Agency (ASI) under contract I/013/12/1. WIND and THEMIS-B/C data were obtained from the NASA-CDAWeb website \url{https://cdaweb.sci.gsfc.nasa.gov/index.html}. Cluster data were obtained from Cluster data repository website \url{https://www.cosmos.esa.int/web/csa}. The list of interplanetary coronal mass ejections, compiled by Ian Richardson and Hilary Cane, can be found at the following website:
\url{www.srl.caltech.edu/ACE/ASC/DATA/level3/icmetable2.htm}. Finally, we like to thank our anonymous Referee for her/his valuable comments.


\bsp
\label{lastpage}
\end{document}